\documentclass[prl,superscriptaddress,showpacs, 12pt]{revtex4}

\usepackage{graphicx}

\begin{document}

\title{Negative Impurity Magnetic Susceptibility and Heat Capacity
in a Kondo Model with Narrow Peaks
in the Local Density of Electron States}
\author{A. K. Zhuravlev\\
Institute of Metal Physics, Ural Division, Russian Academy of Sciences,
 S. Kovalevskoi str. 18, Ekaterinburg, 620041 Russia, E-mail: zhuravlev@imp.uran.ru}


\begin{abstract}
Temperature dependencies of the impurity magnetic susceptibility, entropy, and heat capacity have
been obtained by the method of numerical renormalization group and exact diagonalization for the Kondo
model with peaks in the electron density of states near the Fermi energy (in particular, with logarithmic Van
Hove singularities). It is shown that these quantities can be {\it negative}. A new effect has been predicted (which,
in principle, can be observed experimentally), namely, the decrease in the magnetic susceptibility and heat
capacity of a nonmagnetic sample upon the addition of magnetic impurities into it.
\end{abstract}

\maketitle


The Kondo problem is one of most intriguing and
important problems in the physics of condensed
state. It was initially introduced \cite{Kondo} for explaining
the minimum of resistance in metallic alloys because
of scattering of conduction electrons on a magnetic
impurity and was later extended to other cases. The
Kondo effect is key for explaining the behavior of
heavy-fermion systems \cite{Hewson}, the anomalous electronic
properties of metallic glasses at low temperatures \cite{Cox}, the quantum dots \cite{qd}, and many other correlated
electronic systems.

The practically all results concerning this model
relate to the case of a band that is flat and wide in comparison with the exchange integral $J$. In particular, this
directly concerns the analytical solution obtained in terms of the Bethe ansatz \cite{Wieg}. However, for real
substances the electron density of states can have features
that cannot be assumed to be wide in comparison with $J$ and, therefore, the approximation of a wide flat zone
can be inadequate in them. As we show below, the specific features of the electron structure of surroundings
can manifest themselves in a \emph{qualitatively} different
behavior of the thermodynamic properties of an impurity in such systems.

Let us write down the Hamiltonian of the Kondo model as follows:
\begin{eqnarray}
H_{sd} &=& \sum_{\mathbf{k}\sigma } \varepsilon _{\mathbf{k}}c_{\mathbf{k}%
\sigma }^{\dagger }c_{\mathbf{k}\sigma } - \sum_{\mathbf{kk'}} J_{kk'} \left[S^+c_{\mathbf{k\downarrow}}^{\dagger }
c_{\mathbf{k'}\uparrow} + S^-c_{\mathbf{k\uparrow}}^\dagger c_{\mathbf{k'}\downarrow} +
S_z\left(c_{\mathbf{k\uparrow}}^\dagger c_{\mathbf{k'}\uparrow} - c_{\mathbf{k\downarrow}}^\dagger
c_{\mathbf{k'}\downarrow}\right)\right] \label{Ham_Kondo}
\end{eqnarray}

As usually \cite{Hewson}, let us write down $J_{kk'}=\alpha_k\alpha_{k'}J$, where $\sum_{\mathbf{k}}|\alpha_k|^2=1$, then extract the expression for the local density of states $\rho(\omega)=\sum_{\mathbf{k}}|\alpha_k|^2\delta(\omega-\varepsilon
_{\mathbf{k}})$, and study
the influence of its features on the properties of the
impurity.

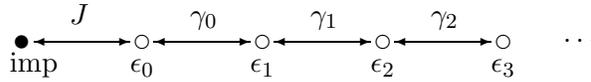
\begin{figure}[ht]
\begin{center}
\unitlength 0.8cm
\begin{picture}(14,2)

\put(1,1){\circle*{0.2}}
\put(3,1){\circle{0.2}}
\put(5,1){\circle{0.2}}
\put(7,1){\circle{0.2}}
\put(9,1){\circle{0.2}}

\put(2,1){\vector(1,0){0.8}}
\put(2,1){\vector(-1,0){0.8}}
\put(4,1){\vector(1,0){0.8}}
\put(4,1){\vector(-1,0){0.8}}
\put(6,1){\vector(1,0){0.8}}
\put(6,1){\vector(-1,0){0.8}}
\put(8,1){\vector(1,0){0.8}}
\put(8,1){\vector(-1,0){0.8}}

\put(0.8,0.5){imp}
\put(1.8,1.3){$J$}
\put(2.8,0.5){$\epsilon_0$}
\put(3.8,1.3){$\gamma_{0}$}
\put(4.8,0.5){$\epsilon_1$}
\put(5.8,1.3){$\gamma_{1}$}
\put(6.8,0.5){$\epsilon_2$}
\put(7.8,1.3){$\gamma_{2}$}
\put(8.8,0.5){$\epsilon_3$}

\put(10,1.0){$\dots$}
\end{picture}
\end{center}
\caption{Representation of the Kondo model in the form of a
semiinfinite Wilson chain.} \label{Fig_chain}
\end{figure}

Following Wilson \cite{Wilson} we use a unitary transformation to pass from the operators $c_{\mathbf{k}}$ to the operators $f_n$:
\begin{eqnarray}
H_{sd}&=& -J\left[S^+f_{0\downarrow}^{\dagger }f_{0\uparrow} + S^-f_{0\uparrow}^\dagger f_{0\downarrow}
 + S_z\left(f_{0\uparrow}^\dagger f_{0\uparrow} - f_{0\downarrow}^\dagger f_{0\downarrow}\right)\right]   \nonumber \\
&+& \sum_{\sigma n=0}^\infty \left[
               \epsilon_n f_{n\sigma}^\dagger f_{n\sigma}
               + \gamma_n \left( f_{n\sigma}^\dagger f_{n+1\sigma}
                  + f_{n+1\sigma}^\dagger f_{n\sigma}\right)\right] \ ,
\label{eq:H_chain}
\end{eqnarray}
and reduce (\ref{Ham_Kondo}) to a semiinfinite chain depicted in Fig.\ref{Fig_chain}, which is called the ``Wilson chain''.

\section{ONE-LEVEL BAND}

Let us begin consideration from a very simple case,
where the band consists of a single level with an energy $\epsilon_0=0$, which corresponds to the case of $\gamma_{0}\rightarrow 0$ in
Fig.\ref{Fig_chain}. This problem is easily solved analytically, and we obtain for the partition function of the entire system $Z_\mathrm{tot}$ and for the single level of a free electron $Z_\mathrm{free}$:
\begin{eqnarray}
Z_\mathrm{tot} &=& 4 + 3e^{\beta J/2} + e^{-3\beta J/2} \ , \\
Z_\mathrm{free} &=& 4 \nonumber \ ,
\end{eqnarray}
where $\beta=1/T$, and $T$ is the temperature.

The impurity magnetic susceptibility $\chi_\mathrm{imp}$ is defined
as a difference in the susceptibilities of the entire system, $\chi_\mathrm{tot}$, and of the system from which the impurity is removed, $\chi_\mathrm{free}$:
\begin{equation}
\label{chi_imp} \chi_\mathrm{imp}=\chi_\mathrm{tot}-\chi_\mathrm{free} \ .
\end{equation}
In our case the total magnetic susceptibility of the entire
system (we assumed that $g\mu_{\rm B}=1$ and $k_{\rm B}=1$) is
\begin{equation}
T\chi_{\mathrm{tot}} = \frac{1}{Z_\mathrm{tot}}\left[1+2e^{\beta J/2}\right] \ ,
\end{equation}
and the susceptibility of free itinerant electrons is
\begin{equation} \label{1free}
T\chi_{\mathrm{free}} = \frac1{Z_\mathrm{free}}\frac{1}{2} = \frac{1}{8}.
\end{equation}

\begin{figure}[htbp]
\includegraphics[width=3.3in, angle=0]{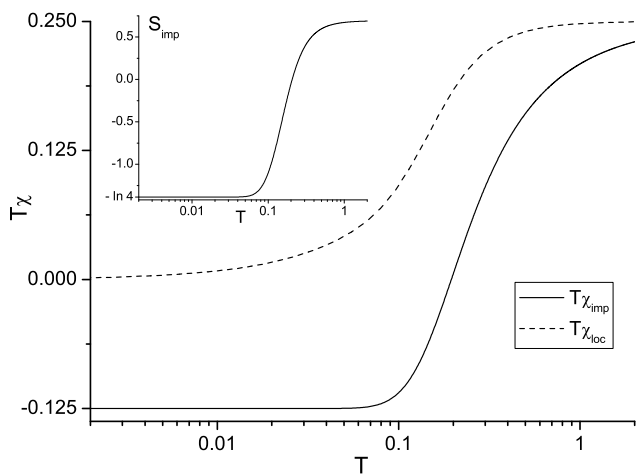}
\caption{Magnetic susceptibilities and the impurity entropy
for a band consisting of a single level; $J=-0.3$.} \label{Fig_Thi1}
\end{figure}

In the case of a negative antiferromagnetic exchange $J$ (at which the Kondo effect does take place), the quantity
$T\chi_{\mathrm{imp}}$ intersects the abscissa axis with decreasing temperature and approaches $-1/8$ as $T \rightarrow 0$
(Fig. \ref{Fig_Thi1}). This result is understandable: at low temperatures, there is formed a singlet with a zero
susceptibility in the spin–electronic system; subtracting the electronic susceptibility (\ref{1free}) from it, we obtain
$-1/8$. The temperature enters into all formulas only in the combination $J\beta$, therefore, any characteristic
temperature will here depend on $J$ according to a linear law.

Apart from the susceptibility (\ref{chi_imp}), the so-called local
magnetic susceptibility $\chi_\mathrm{loc}$ is frequently introduced as
well:
\begin{equation}
  \label{ChiLocal}
    \chi_\mathrm{loc}(T) = \int\limits_0^\beta\langle S_z(\tau)S_z\rangle d\tau \ ,
\end{equation}
where $\langle S_z(\tau)S_z\rangle = \frac{1}{Z}{\rm Tr}\left[e^{-\beta H}e^{\tau  H}S_ze^{-\tau H}S_z\right]$.
This is the susceptibility of a \emph{single} impurity in a magnetic field that
acts locally, only on this impurity; its magnitude, therefore, can hardly be measured experimentally, in contrast to $\chi_\mathrm{imp}$. We thus obtain
\begin{eqnarray}
T\chi_{\mathrm{loc}} = \frac{1}{Z_\mathrm{tot}}
\left[1+\frac{1}{2}e^{\beta J/2} 
+\frac{e^{\beta J/2}-e^{-3\beta J/2}}{4J\beta}\right]
\end{eqnarray}

As we see (Fig. \ref{Fig_Thi1}) $\chi_{\mathrm{imp}}$ and $\chi_{\mathrm{loc}}$ behave quite differently. In principle, this possibility was mentioned in \cite{Santoro}, where the reason for the difference is connected exactly
with the inconstancy of $\rho(\omega)$ and where is asserted that
the difference between them disappears for a flat band
of half-width $D$ in the limit of $D\rightarrow\infty$. However, to the
best of our knowledge, the fact that $\chi_{\mathrm{imp}}$ can become
negative never has been indicated.

For the impurity entropy, we have
\begin{eqnarray}
S_\mathrm{imp} \equiv S_\mathrm{tot}-S_\mathrm{free} = \frac1{TZ_\mathrm{tot}}[3(- J/2) e^{\beta(J/2)} +
(3J/2)e^{-\beta(3J/2)}] + \ln\frac{Z_\mathrm{tot}}{Z_\mathrm{free}} \label{entropy}
\end{eqnarray}
The impurity entropy grows monotonically from $-\ln4$ (at $T=0$) to $\ln2$ (at $T=\infty$) (see inset in Fig.\ref{Fig_Thi1}). This can
easily be understood even without calculations. Indeed,
at a zero temperature the entropy is equal to the logarithm of the multiplicity of the degeneracy of the
ground state, while at high temperatures it is equal to
the logarithm of the number of all states. At $T=0$, the
entropy of the entire system is $S_\mathrm{tot}=\ln
1=0$ (nondegenerate ground state) and the entropy of the electronic part
is $S_\mathrm{free} = \ln4$ (four states with an identical energy); therefore, $S_\mathrm{imp}(T=0)=-\ln4$. At $T=\infty$, the entropy of
the entire system is $S_\mathrm{tot}=\ln 8$ (in the system consisting
of an impurity and an electronic site, there are eight
states); therefore, $S_\mathrm{imp}(T=\infty)=\ln8-\ln4=\ln2$.

As we see, this simple system reveals features that
are absent in the well-investigated case of a wide flat
band, namely, a \emph{negative} impurity susceptibility and a
\emph{negative} entropy. In the general case, we can speak of
two limiting cases: wide flat band and infinitely narrow
infinitely high band (single-level band). All the other
possible bands are ``intermediate'' between these two,
and it is quite probable that some features of the single-level system will be preserved also for these
``intermediate'' cases, which will be considered below in some detail.

\section{BAND OF A FINITE WIDTH}

Let us examine a symmetrical band of finite width without singularities in an interval $[-D, D]$.

\subsection{Analytical Results for $J=-\infty$}

In the case of $J=-\infty$, the Kondo model becomes considerably simpler for calculations. When calculating the eigenenergies of the impurity term
in  (\ref{eq:H_chain}), it proves that the lowest energy state is the singlet, and all excited states are separated from it by a gap of about
$|J|$ and, therefore, they can be neglected. Since the matrix elements of the operators $f_0$ and $f_0^\dagger$ on this singlet are
equal to zero, then the neglect of these excited states reduces simply to the elimination of members
with $f_0^\dagger f_1$ and $f_1^\dagger f_0$ from Hamiltonian (\ref{eq:H_chain}) and the replacement of the impurity member
by its eigenvalue for the singlet state:
\begin{eqnarray}
\label{eq:H_chain_Jinf} H_{sd}(J=-\infty)= \texttt{const} + \sum_{\sigma n=1}^\infty \left[
               \epsilon_n f_{n\sigma}^\dagger f_{n\sigma}
               + \gamma_n \left( f_{n\sigma}^\dagger f_{n+1\sigma}
                  + f_{n+1\sigma}^\dagger f_{n\sigma}\right)\right] \ ,
\end{eqnarray}
so that the problem is reduced to the problem of noninteracting particles, whose solution is much simpler. For
example, to calculate $\chi_\mathrm{imp}$, it is necessary merely subtract the susceptibility $\chi_\mathrm{free}$ of the chain beginning from
the site 0 from the susceptibility of the chain that begins
from the site 1.

It is possible to analytically solve the case of a semielliptical local density of states $\rho(\omega)=2(D^2-\omega^2)^{1/2}/\pi D^2$.
It is shown in \cite{Hewson} that in this case all the coefficients are $\gamma_n=D/2$ and $\epsilon_n=0$. The energies of this
chain of $N$ sites are equal to $E_n=D\cos[\pi n/(N+1)]$, where $n=1, 2, ... , N$. We calculate the total density of
states $\mathcal{D} = dn/dE$ at $n\approx N/2$ (i.e., at the Fermi level $E_F$) and obtain $\mathcal{D}=(N+1)/\pi D$. The Pauli susceptibility at
the zero temperature is equal to $\chi=\frac{1}{2}\mathcal{D}(E_F)$. Therefore,
\begin{equation}
\chi_\mathrm{imp}(J=-\infty, T=0) = \frac{1}{2}\frac{N}{\pi D} - \frac{1}{2}\frac{N+1}{\pi D} = -\frac{1}{2\pi D}.
\end{equation}
If we now select $D=4/\pi$, so that the height at the Fermi level be $\rho(0)=1/2$, then we obtain $\chi_\mathrm{imp}=-1/8$.

As far as the impurity entropy and heat capacity are
concerned, by using a low-temperature expansion for
the ideal Fermi gas known from the textbooks, we
obtain
\begin{equation}
S_\mathrm{imp}(J=-\infty, T\ll1) = \frac{2\pi N}{3D}T - \frac{2\pi(N+1)}{3D}T = -\frac{2\pi}{3D}T.
\end{equation}
and, according to the well-known formula $C=T\frac{dS}{dT}$, the same for the impurity heat capacity. Thus, at low
temperatures, we have $S_\mathrm{imp} = \gamma_\mathrm{imp} T$ and $C_\mathrm{imp} = \gamma_\mathrm{imp} T$, where
$\gamma_\mathrm{imp}=-\frac{2\pi}{3D}$; i.e., the impurity heat capacity $C_\mathrm{imp} \equiv
C_\mathrm{tot}-C_\mathrm{free}$ is \emph{negative}.

An important role in the Kondo problems is played by the Sommerfeld relation, which is here more frequently called the Wilson relation
\begin{eqnarray}
R = \frac{\chi_\mathrm{imp}/\chi_\mathrm{c}}{\gamma_\mathrm{imp}/\gamma_\mathrm{c}} =
\frac{4\pi^2k_B^2}{3(g\mu_B)^2}\frac{\chi_\mathrm{imp}}{\gamma_\mathrm{imp}} \ ,
\end{eqnarray}
where the subscript ``c'' indicates the gas of free electrons. In this case $R=1$ (this is understandable, since
the effective Hamiltonian (\ref{eq:H_chain_Jinf}) is the Hamiltonian of the
ideal gas), while Wilson's calculations for small $J$ gave a value $R\approx2$.

The same can be made semianalytically for the case of a rectangular local density of states: $\rho(\omega)=1/2$ for
$|\omega|<1$. Assuming in formula (\ref{gamma}) (see Appendix) that $\Lambda\rightarrow 1$, we obtain the following coefficients for the
Wilson chain for a flat band:
\begin{equation}
\gamma_n = \frac{n+1}{\sqrt{(2n+1)(2n+3)}}\ , \epsilon_n=0 \ .
\end{equation}
After this, for each length of chain $N$ we can easily numerically determine the electronic spectrum and,
consequently, also the density of states $\mathcal{D}$ and the magnetic susceptibility at $T=0$.
\begin{figure}[htbp]
\includegraphics[width=3.3in, angle=0]{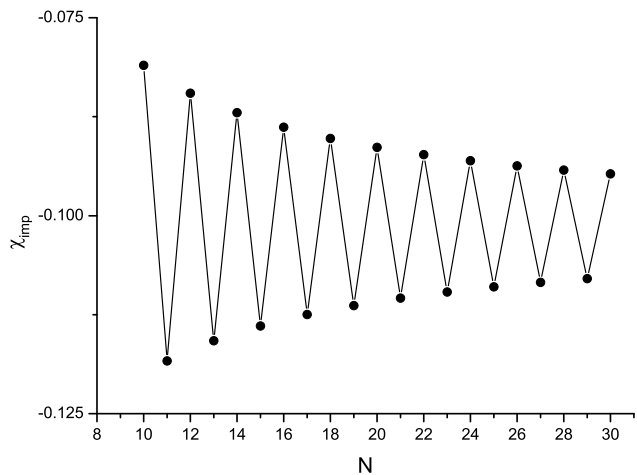}
\caption{$\chi_\mathrm{imp}(T=0)$ for a flat band at $J=-\infty$ as a function of the length of the Wilson chain $N$.}
\label{Fig_FB_Lam1_Jinf}
\end{figure}

With increasing chain length $N$, the magnetic susceptibility $\chi_\mathrm{imp}$ tends to $-1/\pi^2$ (oscillates near this value,
approaching it; see Fig. \ref{Fig_FB_Lam1_Jinf}). Analogously, $\gamma_\mathrm{imp}\rightarrow-4/3$. This differs significantly from the
above-calculated values for the semielliptical local density of states in spite of the fact that at the Fermi level they both have not
only equal values, but also equal values of the derivatives.

\begin{figure}[htbp]
\includegraphics[width=3.3in, angle=0]{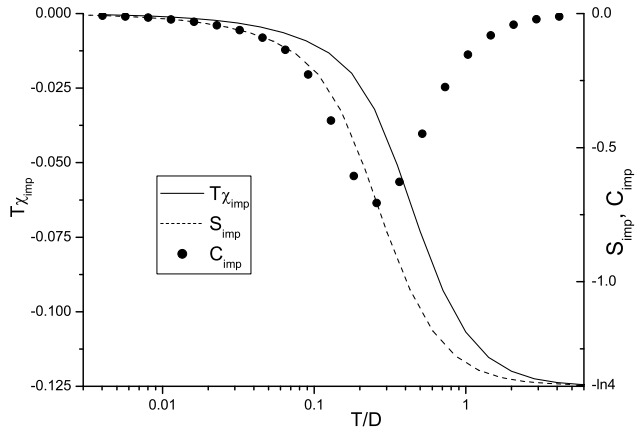}
\caption{$T\chi_\mathrm{imp}$, $S_\mathrm{imp}$, and $C_\mathrm{imp}$ for a flat band of half-width $D$ at $J=-\infty$.}
\label{Fig_FB_Jinf_Tfin}
\end{figure}

Given the entire one-electron spectrum, it is possible to numerically determine $\chi_\mathrm{imp}$, $S_\mathrm{imp}$, and $C_\mathrm{imp}$ for any
temperature. The results obtained are given in Fig. \ref{Fig_FB_Jinf_Tfin}. Note separately that the impurity heat capacity
$C_\mathrm{imp}$ is negative, and has a minimum equal to -0.7 at $T/D\approx0.25$.

Emphasize that all above-mentioned data are the result of strict calculations made without any approximations.

\subsection{NRG Calculations for Finite $J$}

For finite $J$, the calculations were performed by the standard method of numerical renormalization group (NRG) (see Appendix).
Their results are given in Fig. \ref{Fig_ThiDifferJ}.
\begin{figure}[htbp]
\includegraphics[width=3.3in, angle=0]{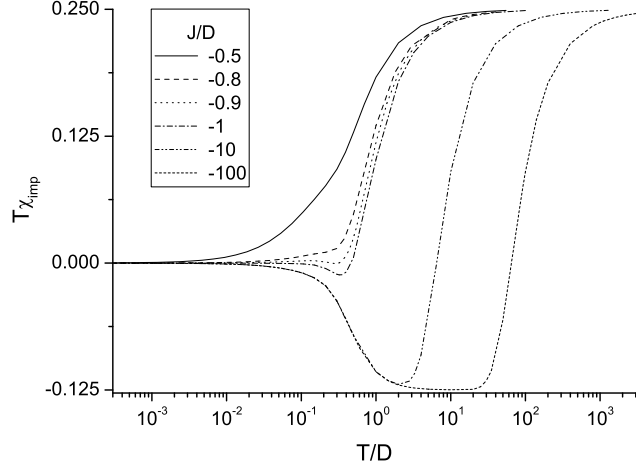}
\caption{Impurity magnetic susceptibility at different $J$.} \label{Fig_ThiDifferJ}
\end{figure}
The local susceptibility $\chi_\mathrm{loc}$ proved to be nonnegative in all the cases, but for $\chi_\mathrm{imp}$ there again are observed
anomalies.

At $|J|$ that are considerably less than $D$ (e.g., at $J/D=-0.5$ in Fig. \ref{Fig_ThiDifferJ}), we see an ordinary picture familiar from
the Wilson works.

At $|J|\gg D$ (e.g., at $J/D=-100$ in Fig. \ref{Fig_ThiDifferJ}), the picture is complicated. At $T\gg D$ (e.g., at $T/D>10$ in
Fig. \ref{Fig_ThiDifferJ}), the behavior observed is quite similar to the case of one level in the band (cf. Fig. \ref{Fig_Thi1});
this is understandable: $D$ is small, and we have $\rho(\omega)$ in the form of a narrow rectangle, similar to the $\delta$-function;
therefore, the results
must be similar: $\chi_\mathrm{imp}\approx-1/8$ at $T<|J|$ and $\chi_\mathrm{imp}\approx1/4$ at $T\rightarrow\infty$. At $T/D<10$, the behavior of
$\chi_\mathrm{imp}$ is very similar to Fig. \ref{Fig_FB_Jinf_Tfin}. This is also understandable: since in this
case $|J|\gg T,D$, the result should be close to the case where $J=-\infty$.

At $|J|\sim D$, there is a transition region between these two regimes; the low-temperature results for it are presented in the table.
Note that $\chi_\mathrm{imp}$ and $\gamma_\mathrm{imp}$ become zero at different $J$.
\begin{table}[htb]
\caption{Low-temperature values of parameters for a rectangular band of half-width $D$ (the errors of determining given values are
about 1\%)}
\begin{tabular}{|c|c|c|c|}
\hline
 $J/D$ & $\chi_\mathrm{imp}(T=0)$ & $\gamma_\mathrm{imp}$ & $R$ \\ \hline
-100 & -0.093 & -1.22 & 1.003  \\
-2   & -0.077 & -1.08 & 0.938  \\
-1   & -0.005 & -0.5  & 0.13   \\
-0.9 &  0.023 & -0.28 & -1.08   \\
-0.85&  0.034 & -0.22 & -2.03   \\
-0.8 & 0.058 & -0.048 & -15.9 \\
-0.75&  0.090 & 0.170 & 6.96    \\
-0.7 &  0.146 & 0.58  & 3.31   \\
-0.5 &  0.603 & 3.6   & 2.2   \\
-0.1 &  13800 & 89000 & 2.04   \\
\hline
\end{tabular}
\label{Tabl_chiJ}
\end{table}

Why have not negative $\chi_\mathrm{imp}$ and $\gamma_\mathrm{imp}$ been revealed in foregoing calculations? This may be
related to the fact that the procedure of calculating susceptibility used by Wilson was suitable only for the case of small
$J$. Wilson used values $|J|\sim 0.05$, and for the low-temperature susceptibility he proposed the formula \cite{Wilson}
\begin{eqnarray}
\chi_\mathrm{imp}=\frac{(0.1032\pm 0.0005)}{\tilde{D}(J\rho)} \left[\exp(\frac{1}{|2J\rho|} -\frac{1}{2}\ln|2J\rho| -
1.5824|2J\rho| + O(J^2\rho^2))\right]
\end{eqnarray}
(where $\tilde{D}(J\rho)\sim D$), which seems to be invalid at large $|J|$. From Fig.~\ref{Fig_ThiDifferJ} it is also seen that the universality
(retention of the form of the curve $T\chi_\mathrm{imp}(T)$ at different {\it J}) discovered by Wilson takes place only for $|J/D|\ll 1$.

\begin{figure}[htbp]
\includegraphics[width=3.3in, angle=0]{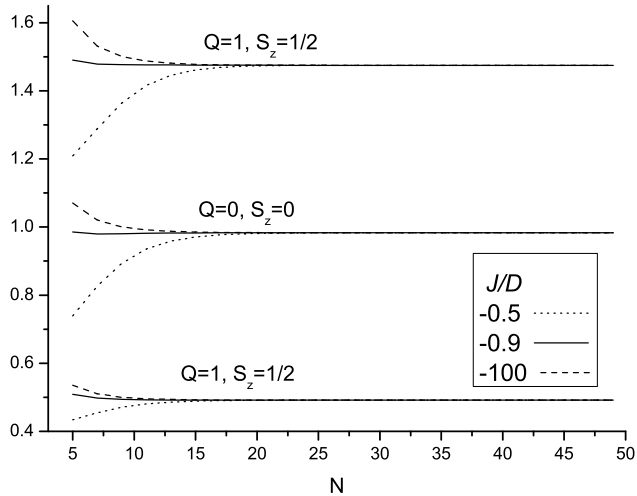}
\caption{The lowest energies multiplied by $\Lambda^{N/2}$ ($\Lambda=2$) for the
$N$th step of the numerical renormalization group, counted
off from the energy of the ground state with $Q=S_z=0$ ($Q$ is
the total charge, i.e., the difference between the numbers of
electrons and holes; $S_z$ is the $z$ projection of the total spin).} \label{Fig_flow}
\end{figure}

The negative values of $\chi_\mathrm{imp}$ and $\gamma_\mathrm{imp}$ appear at $J/D\sim-1$. This it is explained by a change in the behavior of
energy levels (Fig. \ref{Fig_flow}). The greater the magnitude of $|J|$, the more rarefied the energy spectrum and, consequently, the
less the total susceptibility $\chi_\mathrm{tot}$ and the total entropy $S_\mathrm{tot}$. At some, generally speaking, unequal critical values
of the exchange integral $J$, they will become less than the independent-of-$J$ $\chi_\mathrm{free}$ and $S_\mathrm{free}$, respectively.

\subsection{Piecewise-Constant $\rho(\omega)$}

Let us now turn to a more realistic model and examine the idealized case of a piecewise-constant local density of states with a rise in the center:
\begin{equation}
\rho(\omega) = \left\{
\begin{array}{ll}
{H} & \mbox{for $|\omega|<w$,} \\
&  \\
{\frac{1/2-Hw}{1-w}} & \mbox{for $w<|\omega|<1$.}
\end{array}
\right.
\end{equation}

\begin{figure}[htbp]
\includegraphics[width=3.3in, angle=0]{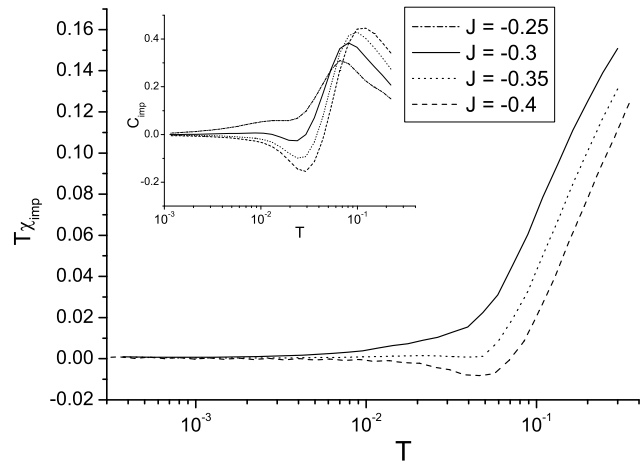}
\caption{$T\chi_\mathrm{imp}$ and $C_\mathrm{imp}$ for a piecewise-constant density of
states: the height at the Fermi level $H=1.0$; the half-width of the peak $w=0.13168$.} \label{Fig_RectPeak}
\end{figure}

The results obtained (Fig. \ref{Fig_RectPeak}) show that negative heat capacity and susceptibility can occur here as well, and
even at smaller values of
$|J|$ in comparison with the flat zone. The general tendency revealed is as follows: the
narrower and the higher the central peak, the smaller the values of $|J|$ at which the negative $\chi_\mathrm{imp}$ and $C_\mathrm{imp}$ appear.

\begin{figure}[htbp]
\includegraphics[width=3.3in, angle=0]{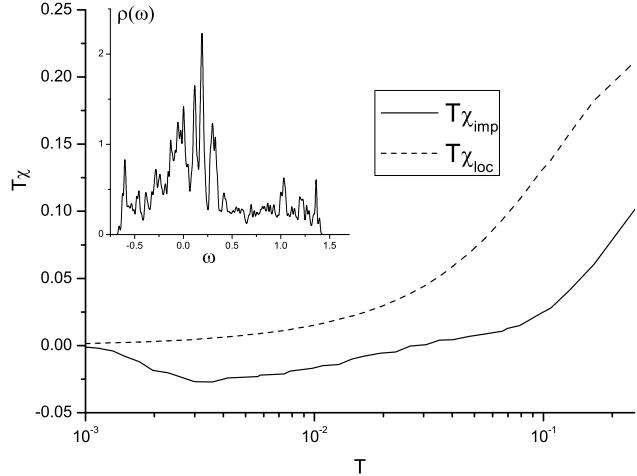}
\caption{Magnetic susceptibilities for an impurity on a disordered two-dimensional lattice; $J=-0.35$.} \label{Fig_point13}
\end{figure}

And finally, quite a realistic case. The inset in Fig. \ref{Fig_point13} depicts the local density of states $\rho(\omega)$ for one of the sites of
the two-dimensional square disordered Anderson lattice taken from \cite{ZhurKett}. The calculation of the magnetic susceptibility of an impurity
located at this lattice site again shows the presence of a negative $T\chi_\mathrm{imp}$.

\section{LOGARITHMIC SINGULARITY IN THE DENSITY OF STATES}

As is well-known, the density of states $\rho(\omega)$ in two-dimensional systems has a Van Hove logarithmic feature. For a two-dimensional square
lattice, we have
\begin{eqnarray}
\rho(\omega) = \frac{2}{\pi^2} K(\sqrt{1-\omega^2})\ , (|\omega|\le 1) \ ,
\end{eqnarray}
where $K$ is the complete elliptical integral of the first kind, which at small $\omega$ behaves as $K(\sqrt{1-\omega^2}) \approx \ln\frac{4}{\omega}$.
It is evident that for this logarithmic feature the characteristic width of the central peak is equal to zero; therefore, in the light of what was said
in the foregoing section it is possible to expect that the negative values of $\chi_\mathrm{imp}$ and $S_\mathrm{imp}$ will appear at \emph{any}
finite $J$.

\begin{figure}[htbp]
\includegraphics[width=3.3in, angle=0]{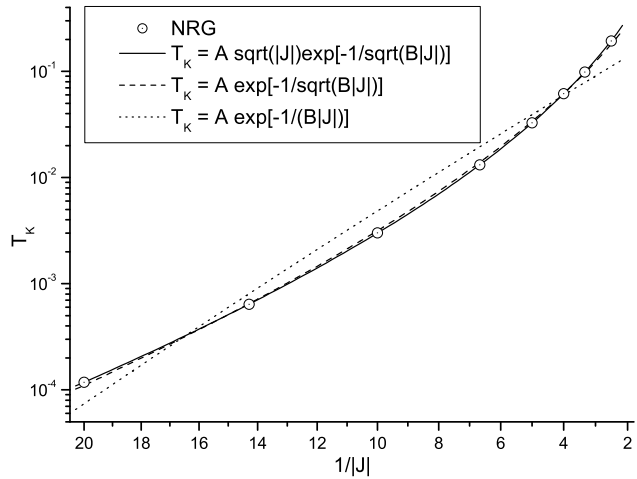}
\caption{Kondo temperature for a two-dimensional square lattice (interpolation by the least-squares method): $A$ and $B$ are
the variable parameters.} \label{Fig_T_K_fit_2D}
\end{figure}

The Kondo model with a logarithmic feature in the density of states was studied in \cite{Gogolin} using the perturbation theory in the parquet
approximation. In that work, the following estimate was given for the Kondo temperature:
\begin{equation}
\label{TK_log} T_K \sim e^{-1/\sqrt{|4\nu_0J|}} \ ,
\end{equation}
where $\nu_0$ is on the order of the inverse bandwidth. This qualitatively differs from the widely known estimate of $T_K$ for the flat band
\begin{equation}
\label{TK_usual} T_K \sim e^{-1/|2J\rho|} \ .
\end{equation}
We verified this assertion by NRG calculations, computing $T_K$ via the formula $T_K\chi_\mathrm{imp}(T_K)=0.0701$ that is
standard in this method \cite{Wilson}. Then, we performed interpolation using the least-squares method with different
trial functions (Fig. \ref{Fig_T_K_fit_2D}), which showed that the results
much better correspond to (\ref{TK_log}) than to (\ref{TK_usual}). And even
more precise is the formula with a root correction:
\begin{equation}
T_K \approx A\sqrt{|J|}e^{-1/\sqrt{B|J|}} \ , \end{equation}
with parameters $A\approx10.08$ and $B\approx0.206$.

\begin{figure}[htbp]
\includegraphics[width=3.3in, angle=0]{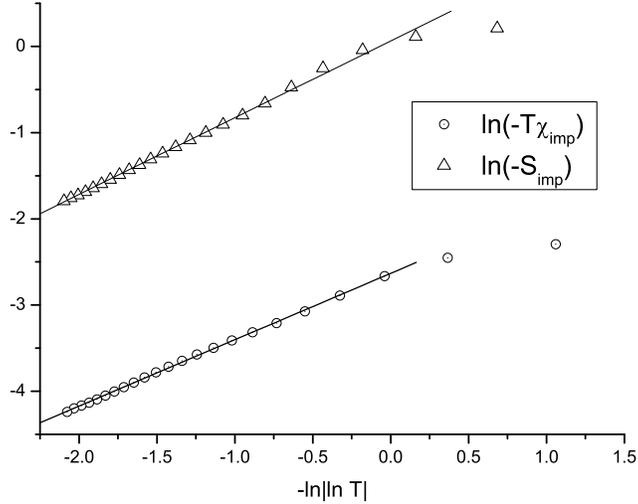}
\caption{Low-temperature behavior in the case of $J=-\infty$ } \label{Fig_log_lowT}
\end{figure}

\begin{figure}[htbp]
\includegraphics[width=3.3in, angle=0]{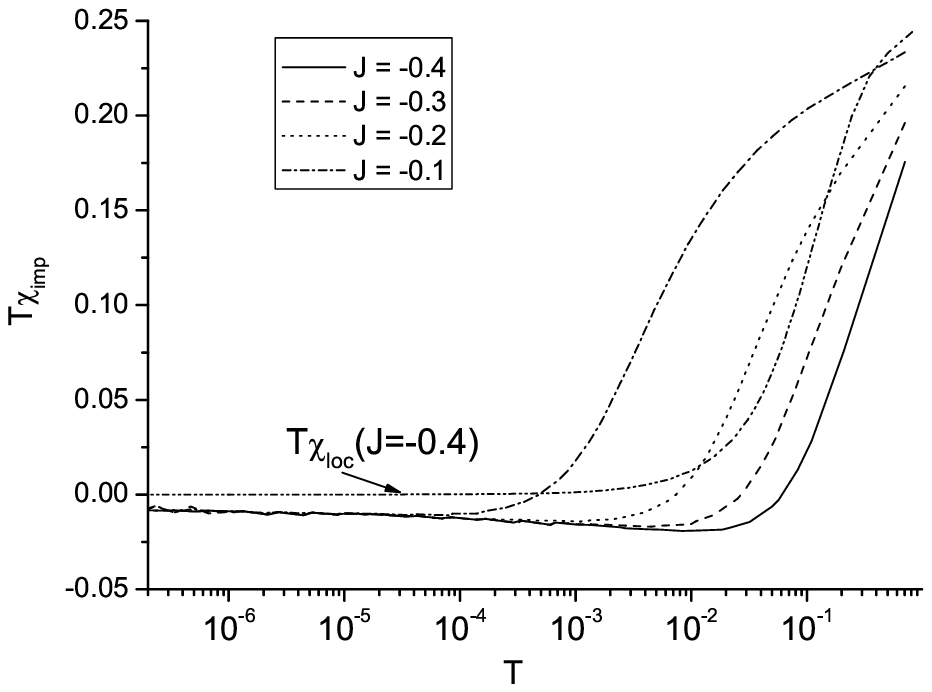}
\caption{$T\chi_\mathrm{{imp}}$ and $T\chi_\mathrm{{loc}}$ for a square lattice.} \label{Fig_Thi_2D}
\end{figure}

Using the procedure described in the preceding section, it is possible to obtain strict results for the case $J=-\infty$. After making a
suitable replacement of variables, it was possible to construct simple linear graphs (Fig. \ref{Fig_log_lowT}),
from which it is easy to determine the low-temperature
behavior of $\chi_\mathrm{imp}$ and of $S_\mathrm{imp}$:
\begin{eqnarray}
 \chi_\mathrm{imp} = -\frac{A}{T|\ln T|^\alpha} \ , (A\approx0.072, \alpha\approx0.77)
 \label{formula} \\
 S_\mathrm{imp} = -\frac{B}{|\ln T|^\delta} \ , (B\approx1.065, \delta\approx0.89).
 \label{formulaS}
\end{eqnarray}
From (\ref{formulaS}), we obtain for the heat capacity
\begin{equation}
 C_\mathrm{imp} = -\frac{B\delta}{|\ln T|^{\delta+1}} .
 \label{formulaC}
\end{equation}
The NRG calculations carried out for finite $J$ showed (Fig. \ref{Fig_Thi_2D}) that nothing uncommon is observed in
the behavior of $T\chi_\mathrm{loc}$. However, $T\chi_\mathrm{imp}$ first intersects the abscissa axis, then begins to grow very slowly,
approaching zero from below. Its low-temperature behavior is independent of $J$ and coincides with that
calculated earlier for the case of $J=-\infty$; the latter is explained by the fact that at sufficiently low temperatures any value of $|J|$
manyfold exceeds both the temperature and the width of the infinitely thin logarithmic peak of $\rho(\omega)$.

\begin{figure}[htbp]
\includegraphics[width=3.3in, angle=0]{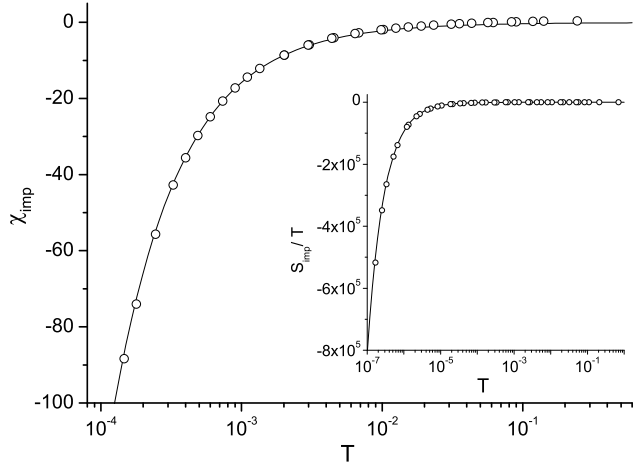}
\caption{Impurity susceptibility and impurity entropy on a square lattice, $J=-0.4$} \label{Fig_2D_hi}
\end{figure}

Thus, a new result is obtained: with decreasing temperature, the impurity susceptibility $\chi_\mathrm{imp}$ and the impurity entropy
$S_\mathrm{imp}/T$ tend to $-\infty$ by the law $-1/T|\ln T|^{\alpha(\delta)}$ (Fig. \ref{Fig_2D_hi}) rather than to $+\infty$, as in a
conventional paramagnet, or to a constant, as in the conventional Kondo regime. In this case, the heat capacity $C_\mathrm{imp}$ is negative
and approaches zero according to a logarithmic law rather than to a linear law.

\section{CONCLUSIONS}

In conclusion, let us emphasize the basic result of
this work: there was predicted a new effect (which, in
principle, can be observed in experiment), namely, a
decrease in the magnetic susceptibility and in the heat
capacity of a nonmagnetic sample upon the addition to
it of magnetic impurities. It must manifest itself in samples in which there are exist sharp peaks in the density
of one-electron states near the Fermi energy, whose
characteristic widths do not exceed the magnitude $J$ of
the exchange interaction of an electron with an impurity. The experimentally observable result must be
especially large for the magnetic susceptibility in samples with a logarithmic feature in the density of states at
the Fermi level.

\section{APPENDIX. Details of the Method of Numerical
Renormalization Group (NRG)}

Here, we briefly describe some moments important
for understanding the essence of the NRG method (this
method has been described in detail in  \cite{Wilson}, \cite{Pruschke}), as well as
some new details that were proposed by the author of this work.

At first, the impurity model with a Hamiltonian of type (\ref{Ham_Kondo}) is reduced to a semiinfinite chain with a Hamiltonian of type
(\ref{eq:H_chain}). The logarithmic discretization and
constructing a Wilson chain are conducted numerically
(except for the calculations of the flat band, which can
be made analytically).

The renormalization-group procedure starts from
the solution of the isolated-impurity problem (sites ``imp'' and $\epsilon_0$ in Fig.~\ref{Fig_chain}). At the first step, there is added
a first conducting electronic site $\epsilon_1$, and there is constructed and diagonalized a Hamiltonian matrix on this
Hilbert space (with a fourfold higher dimensionality). This procedure is repeated multiply. However, since the
dimensionality of Hilbert space grows as $4^N$ ($N$ is the order number of an iteration), it is not possible to store
all eigenstates during the computation. Therefore, it is
necessary to preserve after each iteration only states
with the lowest energies. For the model in question, if
we limit ourselves by a certain maximum number of
stored states (determined by the possibilities of the
computer at hand), it is necessary, beginning from a
certain iteration, to leave on the order of 1/4 of states at
each step.

Unfortunately, the direct application of this diagram
meets with failure, since the disturbance introduced by
the rejection of the high-lying states proves to be too
large. Wilson found a method of overcoming this difficulty; it reduces to the artificial introduction of an exponential damping of matrix elements $\gamma_n$, which decreases the coupling between the preserved and rejected states,
thus decreasing the influence of the rejected states. To
this end, Wilson \cite{Wilson} used a logarithmic discretization of
the conduction band, i.e., replacement in (\ref{Ham_Kondo}) of an
energy range $\varepsilon _{\mathbf{k}} \in (D\Lambda^{-n-1} , D\Lambda ^{-n}],\
n=0,1,2,...$ by a single level with an energy equal to the average energy
of this interval ($D$ is the half-width of the conduction
band, $\Lambda>1$). As a result, after passage from (\ref{Ham_Kondo}) to (\ref{eq:H_chain}) the jumps will have the required damping:
$\gamma_n\propto \Lambda ^{-n/2}$. For a flat band, Wilson obtained analytically
\begin{equation}
\gamma_n =  \frac{D\left( 1+ \Lambda^{-1} \right) \left(1-\Lambda^{-n-1}\right) }{2\sqrt{1-\Lambda^{-2n-1}}           \sqrt{1-\Lambda^{-2n-3}}}\, \Lambda^{-n/2} \ , \epsilon_n=0 \ .
\label{gamma}
\end{equation}
The algorithm of obtaining the result for the initial nondiscretized band is as follows: to conduct the calculation of the observed impurity
value at a given $\Lambda$, changing the number of states preserved at iteration, and to make sure that the result is independent of it, i.e., it is
reliable for a given $\Lambda$; to make this for different $\Lambda$ (for example, for $\Lambda$ = 3, 2.5, 2, 1.5) and, using those values
for which it was possible to obtain reliable results, to make an extrapolation $\Lambda\rightarrow1$.

Now, we concern the new details of the method that were suggested by the author. Earlier, when numerically passing from (\ref{Ham_Kondo}) to (\ref{eq:H_chain}), the researchers attempted to numerically conduct a unitary transformation from
the operators $c_{\mathbf{k}}$ to the operators $f_n$. For this, it is necessary to solve an infinite set of linear equations, which is
impossible; therefore, for example, in \cite{Chen}, the number
of equations was artificially limited. We instead used a
method (described in \cite{Hewson}) of constructing a Wilson
chain via three-diagonalization according to Lanczos \cite{Parlett}, which consists in iteratively constructing such a
space (the so-called Krylov space) in which the operator
\begin{eqnarray}
H_{\sigma} = \sum_\mathbf{k} \varepsilon _{\mathbf{k}}c_{\mathbf{k} \sigma }^{\dagger }c_{\mathbf{k}\sigma } \label{H_c}
\end{eqnarray}
has a three-diagonal matrix. Starting from the vector $|0\rangle = f_{0\sigma}^\dagger |\texttt{vac}\rangle$ (where $f_{0\sigma}^\dagger = \sum_{\mathbf{k}} \alpha_{k}c_{\mathbf{k}\sigma }^{\dagger }$) and using the formula
\begin{eqnarray}
H_{\sigma}|n\rangle = \gamma_n|n+1\rangle + \epsilon_n|n\rangle + \gamma_n^*|n-1\rangle \label{lanczos}
\end{eqnarray}
we consecutively obtain the coefficients $\gamma_n$ and $\epsilon_n$ for (\ref{eq:H_chain}).

As is indicated in \cite{Wilson}, because of the retention of
only part of the energy spectrum at the $N$th step of the
renormalization group procedure, the thermodynamic
averages should be calculated at a temperature that
depends on $\Lambda$: $T_N = \Lambda^{-N/2}T_0$, where the starting temperature $T_0$ is selected more or less arbitrarily. We made an
attempt to select the last value in a regular way. In a
chain of finite dimension, the thermodynamic quantities experience even–odd oscillations depending on $N$ (see Fig.~\ref{Fig_FB_Lam1_Jinf}). It was discovered, that the magnitude of
these oscillations depends on the choice of $T_0$ and has a
clearly pronounced minimum; the temperature corresponding to this minimum was selected as starting.

To finally suppress these even–odd oscillations, we used smoothing according to Euler \cite{Hardy}, which reduces
to the following. If there is a certain oscillating sequence $A_n$, we introduce a new sequence $A^{(1)}_n$, whose
members are equal to arithmetic mean of the adjacent members of the initial sequence:
$A^{(1)}_n= (A_n+A_{n+1})/2$. If necessary, the procedure is repeated: $A^{(2)}_n=
(A^{(1)}_n+A^{(1)}_{n+1})/2$. In particular, this was made in the calculation
of $\chi_\mathrm{imp}$. By designating $\chi_{N} \equiv
\chi_\mathrm{imp}(T_N)$, we obtain
\begin{eqnarray}
&\chi^{(1)}(\sqrt{T_NT_{N+1}}) = \frac{1}{2}\chi_{N} + \frac{1}{2}\chi_{N+1}  \nonumber \\
&\chi^{(2)}(T_N) = \frac{1}{4}\chi_{N-1} + \frac{1}{2}\chi_{N} + \frac{1}{4}\chi_{N+1} \\
\nonumber &\chi^{(3)}(\sqrt{T_NT_{N+1}}) = \frac{1}{8}\chi_{N-1} + \frac{3}{8}\chi_{N} + \frac{3}{8}\chi_{N+1}+
\frac{1}{8}\chi_{N+2}
\end{eqnarray}
With the aid of this method, we determined the low-
temperature values of $\chi_\mathrm{imp}$ and $\gamma_\mathrm{imp}$, which are given in
the table \ref{Tabl_chiJ}.

\section{ACKNOWLEDGMENTS}

This work was supported in part by the Grant of the
President of the Russian Federation for the support of
leading scientific schools, no. NSh-1941.2008.2.

\end{document}